# Tunable Fröhlich Polarons in Organic Single-Crystal Transistors

I. N. Hulea[1], S. Fratini[2], H. Xie[1], C.L. Mulder[1], N.N. Iossad[1], G. Rastelli[2,3], S. Ciuchi[3], and A. F. Morpurgo[1]

[1]*Kavli Institute of Nanoscience, Delft University of Technology, Lorentzweg 1, 2628 CJ Delft, The Netherlands*

[2]*Laboratoire d'Etudes des Propriétés Electroniques des Solides, CNRS, BP 166 - 25, Avenue des Martyrs, F-38042 Grenoble Cedex 9, France*

[3]*INFM-CNR SMC and Dipartimento di Fisica, Università dell'Aquila, via Vetoio, I-67010 Coppito-L'Aquila, Italy*

**In organic field effect transistors (FETs), charges move near the surface of an organic semiconductor, at the interface with a dielectric. In the past, the nature of the microscopic motion of charge carriers -that determines the device performance- has been related to the quality of the organic semiconductor. Recently, it has been appreciated that also the nearby dielectric has an unexpectedly strong influence. The mechanisms responsible for this influence are not understood. To investigate these mechanisms we have studied transport through organic single crystal FETs with different gate insulators. We find that the temperature dependence of the mobility evolves from metallic-like to insulating-like with increasing the dielectric constant of the insulator. The phenomenon is accounted for by a two-dimensional Fröhlich polaron model that quantitatively describes our observations and shows that increasing the dielectric polarizability results in a crossover from the weak to the strong polaronic coupling regime.**

The field of plastic electronics has developed at an impressively fast pace over the last ten years, up to the point that the first commercial applications are now starting to appear. For organic transistors, these developments have been made possible by a intense research effort, that has mainly focused on low-cost thin-film devices. In order to improve the carrier mobility in the devices, particular attention has been devoted to controlling the chemical purity and structural quality of the organic semiconductor, where the motion of the charge carriers takes place[1]. The fact that also the dielectric properties of the gate insulator have a large influence on the mobility has been discovered only recently[2,3] and has come as a surprise. As gate insulators with different dielectric constant can result in devices exhibiting mobility values differing by more than one order of magnitude, the fundamental understanding of this effect is now crucial for further optimization of FET devices.

Our work relies on temperature-dependent electrical transport measurements through rubrene single-crystal FETs. In contrast to thin-film devices, recent experiments have demonstrated that the influence of disorder on transport in these single-crystal transistors is negligible[4-6] near room-temperature. So far, however, these experiments have not yet given precise indications as to the microscopic nature of charge carriers at organic/dielectric interfaces. Most theories developed to describe transport in organic crystals have considered the possibility that charge carriers behave as Holstein polarons, i.e. quasiparticles formed by a charge carrier bound to a short-range deformation of the molecular crystal[7]. The formation of Fröhlich polarons —quasiparticles consisting of a charge carrier bound to an ionic polarization cloud in the surrounding medium— has received much less attention, because it is not expected to occur in the bulk of organic semiconductors. Nevertheless, in FETs, where charge carriers move at the interface with a polar dielectric the role of Fröhlich polarons needs to be addressed[8].

To address this issue, we have used rubrene single crystal FETs fabricated using different techniques[9], that enabled us to vary the dielectric constant of the gate insulator from 1 to 25 (see Methods section). A total of more than 100 FETs were studied, in which the gate dielectric was vacuum ($\varepsilon=1$), parylene ($\varepsilon=2.9$), $SiO_2$ ($\varepsilon=3.9$), $Si_3N_4$ ($\varepsilon=7.5$), $Al_2O_3$ ($\varepsilon=9.4$), and $Ta_2O_5$ ($\varepsilon=25$). Fig.1 (c) and (d) show two single crystal rubrene devices. Examples of source-drain current versus gate voltage ($I_{SD}$-$V_G$) curves measured at different temperatures are shown in Fig. 1 (e) and (f) (all the measurements presented in this paper have been performed along the high-mobility b-direction). From these curves we extract the value of the carrier mobility from the linear regime, and of the threshold voltage $V_{TH}$, by extrapolating the linear part of the $I_{SD}$-$V_G$ curve to zero current. Measurements were performed in between 300 K and 200 K. Below this temperature the difference between the thermal expansion of the organic material and of their supporting substrates often causes device failure due to the formation of crack in the crystals. In the explored temperature range, the observed mobility was weakly or not dependent on gate voltage (see Fig. 2). Only for FETs on $Ta_2O_5$, at carrier densities much higher than those relevant for the present study, the mobility exhibits a rapid decrease with increasing $V_G$. We believe that this may be due to the onset of interaction between electrons that may start to affect the mobility when the density of charge carriers is sufficiently high. This high-density regime is outside the scope of this paper and will be discussed elsewhere.

Figure 3 shows the temperature dependence of the mobility for transistors fabricated with the six different gate insulators, representative of the general behavior of these devices. The mobility tends to decrease with increasing the dielectric constant of the gate insulator[3]. The trend is very systematic, apart for the case of $Si_3N_4$, for which a slightly higher mobility would be expected. We believe that this deviation originates from the formation of a surface layer of $SiO_xN_y$ during the device fabrication, whose dielectric constant was shown to vary from 5 to 34, depending on the stoichiometry[10] (i.e., for $Si_3$

$N_4$ the bulk dielectric constant is not representative of the surface dielectric properties). The most important aspect of our observations is the crossover from "metallic-like" ($d\mu/dT<0$) to "insulating-like" ($d\mu/dT>0$) in the temperature dependence of the mobility that occurs as $\varepsilon$ increases. The effect is large, as the mobility changes by two orders of magnitude at T= 200 K with varying $\varepsilon$ from 1 to 25.

To investigate the dependence of $\mu$ on $\varepsilon$, we have also measured the threshold voltage as a function of $\varepsilon$ and $T$, and found that for all devices $V_{TH}$ varies linearly with $T$ in between 200 and 300 K (see Fig. 4(a)). The shift of $V_{TH}$ with lowering temperature is normally attributed to an increased trapping at impurities[5,11]. Therefore, if the observed influence of the dielectric on the mobility was due to trapping at impurities, we should expect that the threshold voltage shift becomes larger as $\varepsilon$ is increased. In practice, the physically relevant quantity characterizing trapping at the dielectric/organic interface is the shift in the threshold charge defined as $dQ/dT=C_i dV_{TH}/dT$, where $C_i$ is the capacitance per unit area[5]. $dQ/dT$ accounts for the difference in thickness and dielectric constant of the gate insulator and permits to compare different devices (see Fig. 4 (a) and (b)).

It is apparent from Fig. 4 (c) that the shift in threshold charge does not exhibit a systematic trend with increasing $\varepsilon$. This indicates that in single crystal FETs the electrical polarization of the dielectric is not the main mechanism affecting the amount of charge trapping at impurities. Rather, the shift in threshold charge is likely to be due to specific chemical groups present at the surface of the different materials[12]. Interestingly, apart from the case of Ta$_2$O$_5$, the shift in threshold charge is highest for devices fabricated on SiO$_2$, which is one of the dielectrics most commonly used in the characterization of organic thin film transistors. More importantly, the absence of a systematic trend between threshold voltage and ε implies that the dependence of μ on ε does not originate from trapping of charge carriers. This is consistent with recent measurements of Hall effect performed on rubrene single crystal FETs with vacuum, parylene[6] and SiO$_2$[13] gate insulators, in which it was shown that in a range of temperatures between 200 and 300 K the mobility obtained from FET measurements does indeed correspond to the intrinsic (Hall) mobility. We conclude that the effect of the dielectric constant of the gate insulator on the mobility of charge carriers is an intrinsic property of dielectric/organic interfaces.

The identification of an intrinsic dependence of the mobility on the dielectric properties of the gate insulator suggests that the observed phenomenon originates from the interaction of the charge carriers with their polar environment. Such an interaction is well-known in condensed matter physics. In its essence it can be described by a Fröhlich hamiltonian, in which free electrons interact with a dispersionless optical phonon of characteristic frequency $\omega_s$. In common inorganic semiconductors (e.g., Si or GaAs), the effective strength of this interaction is weak due to both the low ionic polarizabilities and the large bandwidths (low band masses) in these materials. As a consequence, the coupling between the carriers and the polar degrees of freedom only causes a small renormalization of the electronic properties. In organic transistors, on the other hand, the bands are narrow (band masses are high) owing to the weak van der Waals bonding of organic crystals, and the use of gate dielectrics with increasing ionic polarizabilities permits to tune the strength of the interaction from the weak to the strong coupling regime. In the latter case, the charge carriers form dielectric polarons. When the polaron radius becomes comparable with the lattice spacing, transport close to room temperature occurs through incoherent hopping between neighboring molecules. It follows that at strong coupling the temperature dependence of the mobility changes from metallic-like to thermally activated. This crossover to the strong coupling regime is the scenario that we invoke to explain the experimental observations.

For strongly coupled Fröhlich polarons of small radius, the expression for the temperature dependent mobility reads[14]:

$$\mu_P(T) = \frac{ea^2}{\hbar} \frac{\omega_s}{T} e^{-\Delta/T} \qquad (1)$$

where $a$ is the hopping length, determined by the distance between neighbouring molecules. The preexponential factor in Eq. (1) is appropriate in the adiabatic regime, i.e. when the phonon frequency is small compared to the bandwidth, which is the case for our experimental system. It is determined to within a numerical factor of order 1 that depends on several microscopic parameters (lattice geometry, interaction with multiple phonon modes[15], polaron size[16]) and that we shall neglect in the following discussion. The activation gap is given in the adiabatic regime by

$$\Delta = \frac{1}{2} E_P - t' \qquad (2)$$

where $E_P$ is the polaron binding energy and $t'$ is a quantity related to, but smaller than, the transfer integral between molecules. The polaron binding energy turns out to be independent of the phonon frequency (see supporting material) and is given by

$$E_P = \frac{a_B}{z} \beta [Ryd] \qquad (3)$$

where $z$ is the distance between the polaron, located in the uppermost one or two molecular layers of the crystal, and the surface of the dielectric. $\beta = \frac{\varepsilon_s - \varepsilon_\infty}{(\kappa + \varepsilon_s)(\kappa + \varepsilon_\infty)}$ is a known parameter that quantifies the ionic polarizability of the interface, expressed in terms of the measured dielectric constants of the gate dielectric ($\varepsilon_s, \varepsilon_\infty$) and of rubrene ($\kappa$)[17]. $a_B$=0.53Å is the Bohr radius.

In order to compare the experimental curves with Eq. 1 we isolate the part of the temperature dependence of the mobility that is caused by the interaction of the charge carriers with the polarizability of the dielectric. In fact, the $\mu(T)$ curves measured experimentally also include a contribution to the temperature dependence due to mechanisms other than Fröhlich polarons, such as the coupling with the molecular vibrations in the organic crystal. The contribution of these other effects can be determined, as they are entirely responsible for the temperature dependence of the mobility measured in FET in which the gate insulator is vacuum: obviously, for these devices the polarizability of the gate dielectric does not play any role. In practice, we use Matthiessen rule $\frac{1}{\mu(T)} = \frac{1}{\mu_P(T)} + \frac{1}{\mu_R(T)}$ as the simplest scheme to decouple the Fröhlich contribution to the mobility $\mu_P(T)$, where $\mu_R(T)$ is the temperature dependent mobility measured on FETs with vacuum as gate insulator.

Fig. 5 (a) shows the experimental data values of $\mu_p(T)$ obtained from Fig 3 using Matthiessen rule, together with best fits based on Eq. 1. In the fitting procedure $\omega_s$ and $\Delta$ are used as free parameters: the resulting values for the different dielectrics are reported in Table 1. As for $\omega_s$, the values obtained from the fits compare well to the known phonon frequencies of the corresponding dielectrics (see Table 1), which is striking given the simplicity of the model and provides clear evidence for the relevance of Fröhlich polarons. In Fig. 5 (b) we plot the values of $\Delta$ for the different dielectrics versus the known $\beta$ values of the corresponding materials. Apart from the case of $Si_3N_4$, for which the formation of $SiO_xN_y$ makes the surface dielectric properties unknown and the comparison with theory impossible, the data show that a linear relation between $\Delta$ and $\beta$ (Eq. 2 and 3) holds as long as $\beta$ is sufficiently large (i.e., for $Ta_2O_5$, $Al_2O_3$, and $SiO_2$). This is expected, since only for sufficiently large $\beta$ the strong coupling analysis based on Eq. 3 is valid. The slope of $\Delta(\beta)$ plot is determined by the distance $z$ between the polaron and the dielectric surface (see Eq. 1,2) and from the data we obtain a value of $z=6.4$ Å that compares well to the molecular size expected in our transistor geometry. From the intercept at zero $\beta$ we obtain a value for $t'$ between 0 and 20 meV that has the same order of magnitude (but is smaller than) the transfer integral known from recent calculations[18-20]. Finally, we note that the model also explains the correlation between mobility and dielectric constant that we had reported in our earlier work[3]. In fact, although the parameter $\beta$ that appears in the theory depends on both $\varepsilon_s$ and $\varepsilon_\infty$, Table I shows that changing the gate insulators mainly results in a change of $\varepsilon_s$, whereas the difference in the values of $\varepsilon_\infty$ for the different materials is a smaller effect.

The theoretical scenario that we have used works quantitatively because it specifically accounts for the microscopic characteristics of the organic material -lattice periodicity and narrow bandwidth- and of the gate dielectric -ionic polarisability and phonon frequencies- that are all needed to give a consistent description of the experimental data. The very good agreement between the values of $\omega_s$ obtained from the fits and the known phonon frequencies of the corresponding dielectrics provides a direct demonstration of this fact. To further validate this conclusion we have compared the experimental data to predictions based on alternative models and found that they fail to give a consistent interpretation. For instance, if the Fröhlich interaction is described in the usual continuous medium approximation[21], it is not possible to reproduce the thermally activated behavior of the mobility, except at unrealistically large coupling strengths. We have also analyzed the anti-adiabatic regime of the model discussed here, where the pre-exponential factor does not depend on the phonon frequency $\omega_s$, and found that in this regime the results of the analysis are not internally consistent (see supplementary material). Finding that the quality of the experimental data is sufficient to discriminate between different microscopic models gives additional support to the conclusions presented above.

We conclude that Fröhlich polarons are indeed formed in organic field-effect transistors if the gate dielectric is sufficiently polar. This conclusion is important for different reasons. First, it represents a considerable step forward in our basic understanding of transport through organic transistors, as it identifies a microscopic physical process that has a large influence on the device performance. Second, it shows that dielectric/organic interfaces provide an ideal model system for the controlled study of Fröhlich polarons, with tunable coupling. We emphasize that these conclusions have been made possible by the high quality and reproducibility of recently developed organic single crystal FETs. As the study of these devices has only started a few years ago, we expect that future research on these systems will extend our fundamental knowledge of electron transport in organic semiconductors by enabling reliable and systematic measurements of many different quantities, such as electronic bandwidth, contact resistance, low temperature transport. The resulting microscopic understanding will permit to establish a solid scientific basis for the physics of organic semiconductors, which will be crucial for future progress in the rapidly expanding field of plastic electronics.

**Methods**

For the fabrication of the rubrene single-crystal FETs we have used three different, recently developed techniques[9]. FETs with a suspended channel -i.e., in which vacuum acts as gate insulator- were fabricated using PDMS stamps molded on a photoresist mask. For the details of the fabrication procedure we refer the reader to Ref.[22]. The electrical characteristics of our devices perfectly reproduce those of similar rubrene FETs previously reported by Podzorov and co-workers[5]. Transistors with Parylene as gate dielectric were fabricated following the process first discussed in Refs.[23,24]. Both evaporated silver electrodes and manually deposited carbon past source and drain electrodes were used in different transistors investigated. On the remaining gate insulators, devices were fabricated by means of electrostatic bonding of very thin (~1 μm thick) rubrene crystals to substrates with a prefabricated FET circuitry, as described by de Boer et al.[25]. In all cases the substrate was degenerately doped crystalline Silicon, covered by the different insulating layers (thermally grown $SiO_2$, chemical vapor deposited $Si_3N_4$, sputtered $Ta_2O_5$ and $Al_2O_3$). Source and drain contacts were made of gold deposited by electron-beam evaporation.

For electrostatically bonded devices, the substrate surface was exposed to an Oxygen plasma prior to the crystal bonding, which we found to be necessary to ensure device reproducibility with this fabrication method[25]. For $Si_3N_4$ substrates this Oxygen plasma results in oxidation of the material, so that the surface consists of $SiO_xN_y$ with an unknown stechiometry, as discussed in the main text. In order to minimize the effect of the oxidized surface, we have tried to fabricate single crystal FETs using a $Si_3N_4$ gate insulator, without performing the oxygen plasma cleaning step. For these devices, however, we could not observe a sufficiently reproducible behavior, as expected from our previous work on different substrates when no oxygen plasma was employed in the substrate preparation.

The data for the mobility shown in Fig. 2 and 3 of the main text are characteristic of high quality devices fabricated on crystals that were re-grown at least three time (starting from rubrene molecules purchased from Sigma-Aldrich) by means of a vapor phase transport technique commonly used form many organic semiconductors. For the device fabrication we used in all cases needle-shaped crystals aligned along the high-mobility b-direction (typical channel length was between 300 μm and 1 mm while the width was typically between 50 and 150 μm). Measurements of the FET electrical characteristics were taken in the vacuum ($10^{-7}$ mbar) chamber of a flow cryostat, in a two terminal configuration with a long channel length (>300μm) to exclude a possible influence of the contact resistance. The mobility is extracted using the usual formula valid for the linear regime. With all these precautions, the spread in mobility values measured in different samples is rather narrow. At room temperature, as a function of gate dielectric: for vacuum, μ~ 15-20 cm$^2$/Vs; parylene μ~ 8-12 cm$^2$/Vs; $SiO_2$ μ~ 4-7 cm$^2$/Vs; $Si_3N_4$ μ~ 2-3 cm$^2$/Vs; $Al_2O_3$ μ~ 2-4 cm$^2$/Vs; $Ta_2O_5$ μ~ 1-1.5 cm$^2$/Vs). The threshold voltage shift measured in different devices when lowering the temperature was highly reproducible (<20% differences in different devices with the same gate insulator). It is worth noting that, contrary to the threshold voltage shift, the absolute value of the threshold voltage can exhibit rather large variations ( 10 Volts) in different devices, and therefore it does not provide a good measure of charge trapping due to impurities.

**Acknowledgements**

We gratefully acknowledge V. Podzorov for discussions and for letting us use his temperature dependent measurements on FETs with parylene gate dielectric. We thank R.W.I. de Boer and A.F. Stassen for contributing to the initial part of this work. Useful discussions with J. van den Brink are also acknowledged. This work was supported by FOM and by NWO through the Vernieuwingsimpuls 2000 program.

**Captions**

Figure 1 High-quality organic single-crystal transistors. (a) Schematic representation of the single-crystal FETs used in our investigation. Electrostatically induced holes moving at the dielectric/organic interface polarize the gate dielectric as illustrated in (b). For sufficiently large coupling (i.e., dielectric constant) the hole and the induced charge can move together, resulting in the formation of a Fröhlich polaron. Panels (c) and (d) are optical microscope images of actual devices (the color is determined by the microscope settings). Typical examples of the device current versus gate voltage measured at different temperatures (1, T= 295 K; 2, T= 270 K; 3, T= 250 K; 4, T= 230 K; 5, T= 210 K) are shown in panel (e) and (f) for devices with vacuum ($V_{SD}$=5 V) and $Al_2O_3$ ($V_{SD}$=1 V) as gate dielectric, respectively.

Figure 2 Gate voltage dependence of the carrier mobility. The panels show the mobility of devices fabricated on different gate dielectrics, extracted from measurements of the source-drain current in the linear regime. In all cases, the mobility is essentially independent on gate voltage, which enables an easy comparison of different devices. In all panels 1 corresponds to T= 295 K, 2 to T= 270 K, 3 to T= 250 K, 4 to T= 230 K, 5 to T= 200 K, 6 to T= 175 K, 7 to T= 150 K. For devices with $SiO_2$ as gate insulator, the mobility is essentially temperature independent in the range investigated, for all values of gate voltage.

**Figure 3** Temperature dependence of the carrier mobility. For single-crystal rubrene FETs with six different gate dielectrics the temperature dependence exhibits a continuous evolution from metallic-like to insulating-like, as the dielectric constant of the gate insulator is increased. For the case of $Si_3N_4$, caution is needed in interpreting the data, because of the formation at the surface of $SiO_xN_y$ -whose dielectric constant has been shown to vary between 5 and 34 depending on the stechiometry- during the device fabrication. In all cases, the values of mobility reported here have been measured at a gate voltage close to $V_G$= -15 V.

**Figure 4** Dependence of the threshold charge on temperature and dielectric constant of the gate insulator. For all devices, the threshold voltage shifts linearly with temperature in the range investigated, irrespective of the dielectric. This is shown in panel (a) for $SiO_2$, where the temperature dependence of the threshold voltage shift is shown for three devices with different thickness of the gate insulator (closed circles - 100 nm, open circles - 200 nm, and closed diamonds - 500 nm). Panel (b) shows that, contrary to the threshold voltage shift, the shift of the threshold charge with temperature does not depend on the thickness of the gate insulator. The shift in threshold charge is a reproducible quantity determined only by the dielectric/organic interface, which can be used to compare different devices. The comparison of the threshold charge shift with temperature for the devices with the six different gate insulators is shown in panel (c). The absence of a systematic relation between $dQ/dT$ and $\varepsilon$ is apparent.

**Figure 5** Comparison between experimental data and theory. (a) The temperature-dependent mobility of Fröhlich polarons (symbols) is compared to the theoretical expression (continuous lines), using $\omega_s$ and $\Delta$ as fitting parameters. For the different dielectrics the values of $\omega_s$ compare well to the known phonon spectra (see Table 1 ). The dependence of $\Delta$ on the dielectric properties of the gate insulator is compared to the theoretical expression valid in the strong coupling regime in panel (b). For $SiO_2$, $Al_2O_3$, and $Ta_2O_5$, $\Delta$ shows a linear dependence on $\beta$, as expected. For parylene, which has a dielectric constant comparable to that of rubrene, the strong coupling theory breaks down, also as expected. The open green circle represents the case of $Si_3N_4$ assuming a dielectric constant of 7.5. The data point falls on the theoretical curve (full green circle) for $\varepsilon=30$, compatible with the known range of possible dielectric constant values of $SiO_xN_y$ (5 to 34)[10], which is formed at the surface of $Si_3N_4$ during the device fabrication.

|  | $\varepsilon_s$ | $\varepsilon_\infty$ | $\omega_{LO}(cm^{-1})$ | $\beta$ | $\omega_s(cm^{-1})$ | $\Delta$ (meV) |
|---|---|---|---|---|---|---|
| $Ta_2O_5$ | 25 | 4.4 | 200-1000 | 0.099 | 315 | 55 |
| $Al_2O_3$ | 9.4 | 3 | 400-900 | 0.086 | 705 | 46 |
| $SiO_2$ | 3.9 | 2.1 | 400-1240 | 0.051 | 620 | 28 |
| parylene | 2.9 | 2.56 | 500-1800 | 0.010 | 1215 | 28 |

**Table 1**: Physical properties of the different gate dielectrics. The values of the static dielectric constant $\varepsilon_s$ have been obtained from capacitance measurements; the values of $\varepsilon_\infty$ are known from the literature. $\omega_{LO}$ represents the range of phonon frequencies for the bulk of the materials found in the literature (see also references in[20]). The values of parameters $\omega_s$ and $\Delta$ have been extracted from the fits of the temperature dependence of the mobility

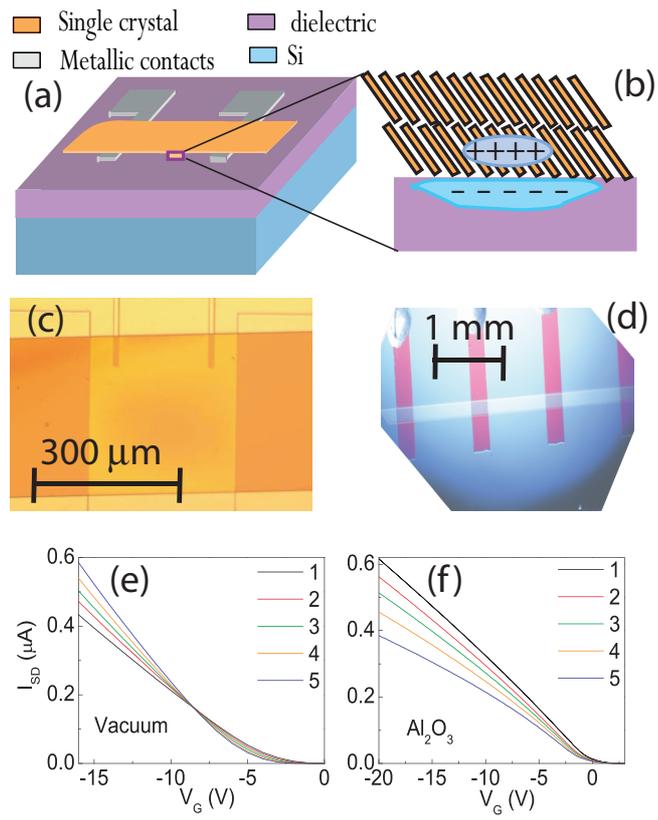

Figure 1, Hulea et al.

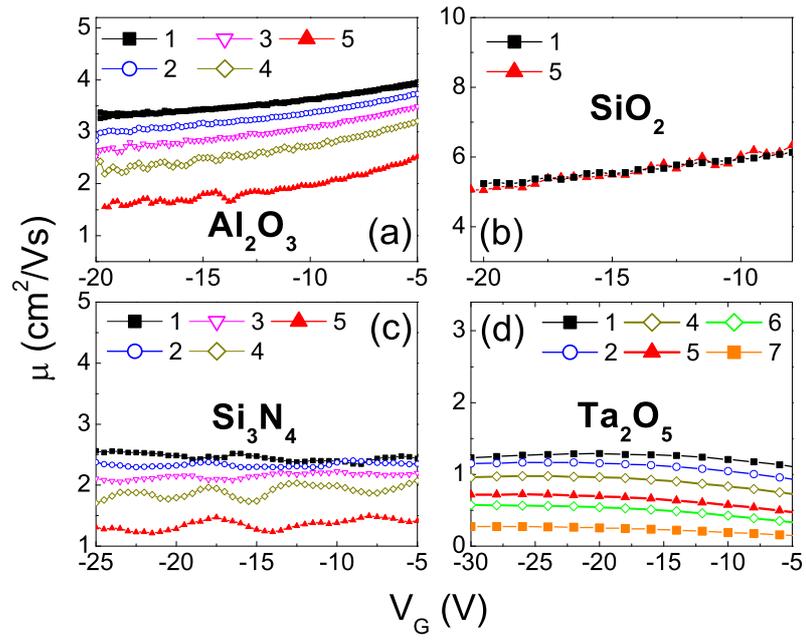

Figure 2, Hulea et al.

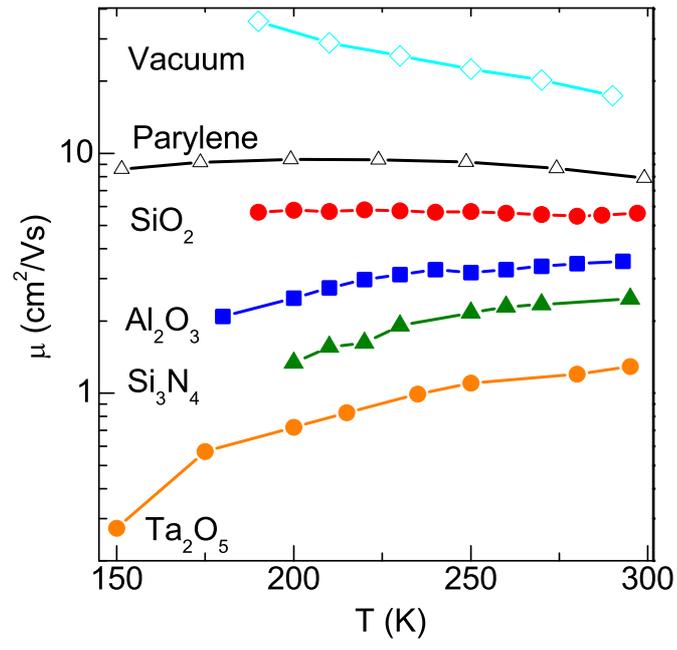

Figure 3, Hulea et al.

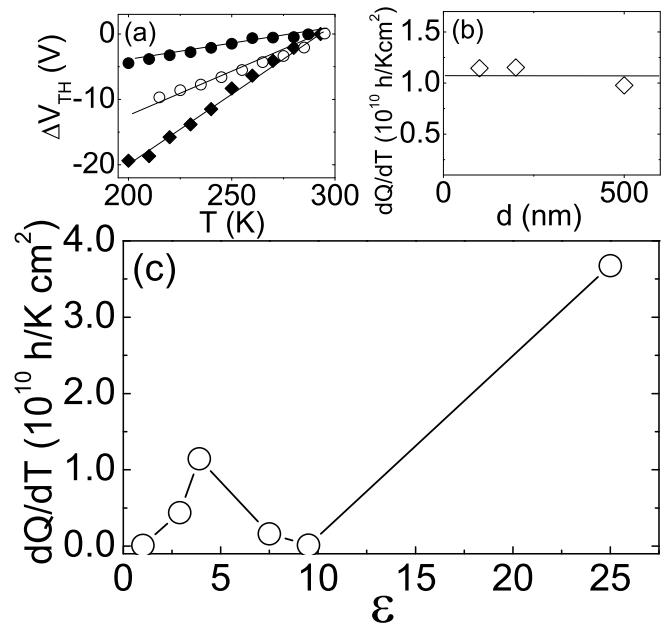

Figure 4, Hulea et al.

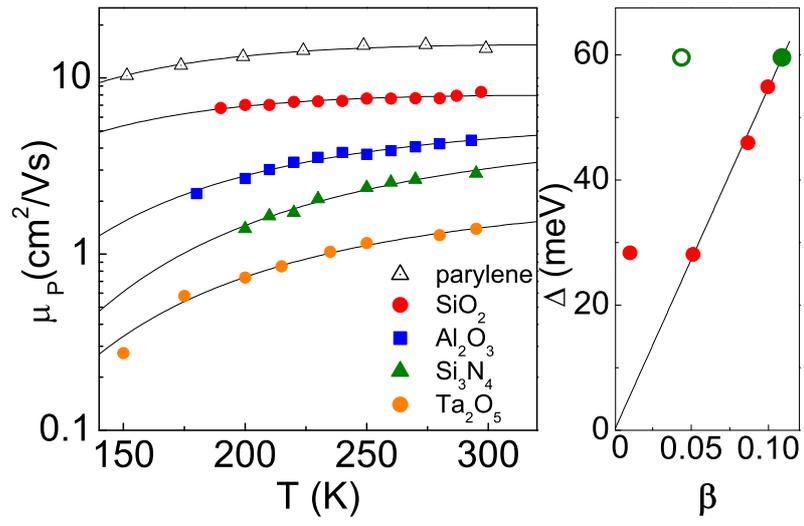

Figure 5, Hulea et al.